\documentclass[american,aps,pra,reprint,floatfix,superscriptaddress,longbibliography]{revtex4-1}
\usepackage[unicode=true,pdfusetitle, bookmarks=true,bookmarksnumbered=false,bookmarksopen=false, breaklinks=false,pdfborder={0 0 0},backref=false,colorlinks=false]{hyperref}
\hypersetup{colorlinks,linkcolor=myurlcolor,citecolor=myurlcolor,urlcolor=myurlcolor}
\usepackage{graphics,epstopdf,graphicx,amsthm,amsmath,amssymb,mathptmx,braket,colortbl,color,bm,framed,mathrsfs}
\usepackage[T1]{fontenc}
\usepackage[up]{subfigure}
\usepackage{tikz}
\usepackage[shortlabels]{enumitem}
\usepackage[normalem]{ulem}
\definecolor{myurlcolor}{rgb}{0,0,0.9}

\usepackage{multirow}
\usepackage{tabularx}

\newcommand{\proj}[1]{| #1\rangle\!\langle #1 |}

\newcommand{\inner}[2]{\langle #1 , #2\rangle}

\DeclareMathOperator{\trace}{Tr}
\newcommand{\Ptr}[2]{\trace_{#1}\Pa{#2}}
\newcommand{\Tr}[1]{\Ptr{}{#1}}
\newcommand{\ptr}[2]{\mathrm{Tr}_{#1}\left\{#2\right\}}

\newcommand{\Pa}[1]{\left[#1\right]}

\newcommand{\norm}[1]{\left\lVert #1 \right\rVert}

\theoremstyle{plain}
\newtheorem{thm}{Theorem}

\newtheorem{prop}[thm]{Proposition}
\newtheorem{cor}[thm]{Corollary}
\newtheorem{Def}[thm]{Definition}

\newcommand*{\myproofname}{Proof}

\def\ot{\otimes}
\def\complex{\mathbb{C}}
\def\real{\mathbb{R}}

\def\CMM{\mathcal{M}}
\def\CNN{\mathcal{N}}

\def\CHH{\mathcal{H}}

\def\sgn{\mathrm{sgn}}

\DeclareMathAlphabet{\mathcal}{OMS}{cmsy}{m}{n}

\makeatother

\begin{document}

\title{
Quantum Entropy and Central Limit Theorem
}

   \author{Kaifeng Bu}
  \email{kfbu@fas.harvard.edu}
 \affiliation{Department of Physics, Harvard University, Cambridge, Massachusetts 02138, USA}

   \author{Weichen Gu}
  \email{weichen.gu@unh.edu}
 \affiliation{Department of Mathematics and Statistics, University of New Hampshire, Durham, New Hampshire  03824, USA}

   \author{Arthur Jaffe}
  \email{jaffe@g.harvard.edu}
 \affiliation{Department of Physics, Harvard University, Cambridge, Massachusetts 02138, USA}
 \affiliation{Department of Mathematics, Harvard University, Cambridge, Massachusetts 02138, USA}

\begin{abstract}

We introduce a framework to study discrete-variable (DV) quantum systems based on qudits. 
It relies on notions of a mean state (MS), a minimal stabilizer-projection state (MSPS), 
and a new convolution. Some interesting consequences are: 
The MS is the closest MSPS to a given state with respect to the relative entropy; the MS is extremal with  respect to the von Neumann entropy, demonstrating a ``maximal entropy principle in DV systems.'' 
We obtain a series of inequalities for quantum entropies and for Fisher information based on convolution, 
giving  a ``second law of thermodynamics for quantum  convolutions.''  We show that the convolution of two stabilizer  states is  a stabilizer state.
We establish a central limit theorem, based on iterating the convolution of a zero-mean quantum state, and show this converges to its MS. The rate of convergence is characterized by  the ``magic gap,'' which we define in terms of the support of the characteristic function of the state.  We elaborate on two examples: the
DV beam splitter and the DV amplifier. 

\end{abstract}

\maketitle

\section{Introduction}
Quantum information and quantum computation come in two forms, continuous-variable (CV) and discrete-variable (DV) systems.  
CV quantum information has been widely used in quantum optics and other settings to deal with continuous degrees of freedom~\cite{SethRMP12}.  
Gaussian states, and processes which can be represented in terms of a Gaussian distribution, are the primary tools used in studying CV quantum information. 
One important property of Gaussian states is their extremality within all CV states, under some constraint on the covariance
matrix~\cite{Holevo99,WolfPRL06,HolevoMutual99,CerfPRL04,Eisert07}.  
Gaussian states also
minimize the output entropy or maximize the achievable rate of communication by Gaussian channels. One sees this using quantum entropy-power inequalities on the convolution of CV states 
\cite{Konig13,KonigPRL13,Konig14,Palma14,PalmaIEE16,Qi2016,Huber17,PalmaPRA15,PalmaIEEE17,PalmaPRL17,PalmaIEEE19}. This statement is a quantum analogue of Shannon's entropy power inequality~\cite{Shannon,Stam,Lieb78}.
These states have both been realized in experiment, and also applied in quantum information tasks, such as quantum teleportation~\cite{Vaidman94,Braunstein98,Tittel98},
quantum-enhanced sensing~\cite{Caves81,Bondurant84,Tan08,Zhuang17},  quantum-key distribution~\cite{Grosshans02} and 
quantum-speed limits~\cite{Becker21}. 

However, 
computational processes with only Gaussian states and processes can be efficiently simulated  on a classical computer~\cite{Bartlett02,Mari12,Veitch_2013}. 
Hence, non-Gaussian states and processes are necessary to implement universal quantum computing~\cite{Lloyd99,BartlettPRL02}. To quantify the non-Guassian nature of a quantum state or process,  the framework of resource theory has been used~\cite{Albarelli18,Takagi18,Chabaud20}.
CV quantum systems have also been considered as a platform to implement quantum computation and realize quantum advantage. Several sampling tasks have been  proposed~\cite{Lund14,Douce17,Hamilton17,Cerf17}, including Gaussian boson sampling, a modification of the original boson sampling proposed by Aaronson and Arkhipov~\cite{aaronson2011computational}. This  has attracted much attention and has been realized experimentally; it is claimed that they display a quantum advantage over classical computers~\cite{Pan20,Pan21,Jonathan22}.

This raises a natural question, ``what states in DV quantum systems play the role of Gaussian states in CV quantum systems?''  Here we focus on   
stabilizer states. They are the common eigenstates of certain abelian subgroups of the qubit Pauli group, and were introduced by Gottesman to study error correction~\cite{Gottesman97}.  
There are several indications that stabilizer states are  the finite-dimensional analogue of Gaussian states in CV quantum systems. For example, the Hudson theorem for CV systems states that the Wigner function of a pure state is nonnegative, if and only if the state is Gaussian \cite{Hudson74,Soto83}.
On the other hand, 
Gross proved in DV systems with the 
local dimension being an odd prime number, that the discrete Wigner function of a pure state  is nonnegative, if and only if the state is a stabilizer~\cite{Gross06}.

From the Gottesman-Knill theorem~\cite{gottesman1998heisenberg}, we infer that stabilizer circuits comprising Clifford unitaries with stabilizer inputs and measurements can be efficiently simulated  on a classical computer. 
In fault-tolerant quantum computation, 
logical Clifford unitaries can be implemented transversally so they are considered to be low-cost. However, 
the
Eastin-Knill theorem~\cite{EastinPRL09} states that there is no quantum error correction code in which any universal gate set can be implemented transversally. Hence, non-stabilizer resources are necessary
to achieve universal quantum computation.  

In recent literature, the property of not being a stabilizer has been called magic. 
To quantify the amount of magic, 
several magic measures have been proposed \cite{Veitch12mag,LeonePRL22,HowardPRL17,BeverlandQST20,SeddonPRXQ21,BravyiPRL16,BravyiPRX16,bravyi2019simulation, Bu19,Bucomplexity22,BuPRA19_stat}, 
and  applied in the classical simulation 
of quantum circuits \cite{SeddonPRXQ21,BravyiPRL16,BravyiPRX16,bravyi2019simulation,Bu19,Bucomplexity22,bu2022classical} and unitary  synthesis \cite{HowardPRL17,BeverlandQST20}. 
Moreover, 
to achieve a quantum advantage for  DV quantum systems, 
several sampling tasks 
have been proposed \cite{jozsa2014classical,koh2015further,bouland2018complexity,boixo2018characterizing,bouland2018onthecomplexity,bremner2010classical,Yoganathan19}.
Some of these proposals have been realized in experiment, which were used to claim a computational advantage over  classical supercomputers \cite{Google19,zucongzi1,Pansci22}.

\subsection{Summary of Main Results}

Little had been known about the extremality of stabilizer states, or their role in the convolution of DV states. 
We propose a framework to study these questions, based on defining a convolution for DV quantum systems. We explain the intuition behind our approach  and state our key results in this paper. 
The complete details and proofs, as well as a theory of the convolution of  quantum channels, appear in an extended, companion work~\cite{BGJ23b}.

Our approach is different from the one in  \cite{Audenaert16, Carlen16}.
Our convolution of states $\rho\boxtimes \sigma$ depends on a chosen Clifford unitary, along with  a partial trace. 
We study our approach, with the  special goal to reveal extermality of stabilizer states in  relation to 
the convolution. This work  includes the following:

\begin{enumerate}[] 
\item{}
We introduce the notion of a mean state (MS), which is the closest state in the set of minimal stabilizer-projection states (MSPS) with respect to the relative entropy in Definition~\ref{def:mean_state}. We prove the  extremality of MSPS: within all quantum states having the same  MS up to Clifford  conjugation, 
the MSPS attains the maximal R\'enyi entropy. 
One implication of the extremality of the MS is that it provides a nontrivial, resource-destroying map in the resource theory of magic; see Corollary~\ref{rem:RD}. 

\item{}
We introduce the notion of the  \textit{magic gap}, which is the difference between the first  and second largest absolute values in the support of the characteristic function in Definition \ref{def:ma_gap}. We prove that the magic gap can serve as a magic measure; it provides  a  lower bound on the number of the non-Clifford gates in the synthesis of the unitary. We formulate these  results in Propositions~\ref{prop:magap} and \ref{prop:gap_syn}.

\item{}
We introduce our convolution $\boxtimes$  in Definition~\ref{def:convo}. A fundamental property is Proposition~\ref{prop:convolutionalstability}, showing that stabilizer states are closed under convolution.  
Convolution also increases the generalized quantum R\'enyi entropy, as stated in Theorem \ref{thm:entropy}. Convolution decreases the Fisher information, as stated in Theorem \ref{thm:fisher}.
We state in Theorem \ref{thm:min_outen} that  the  convolutional channel  achieves minimal output entropy, if and only if the input states  are pure stabilizer states. We study the Holevo channel capacity of the convolutional channel, and  show that the convolutional channel achieves the maximal Holevo capacity if and only if the state is a stabilizer, see Theorem \ref{thm:holv_stab}. 

\item{}
Our convolutional approach includes two important examples, the DV beam splitter and 
the DV amplifier, both of which share a similar structure to their CV counterparts. 
We compare our new DV  results on the beam splitter to the known results for CV quantum systems in \S\ref{subsec:examp}, Table~\ref{tab:sum_B}. We also compare CV and DV cases for the amplifier in \S\ref{subsec:examp},  Table~\ref{tab:sum_S}. 

\item{}
We establish a quantum central limit theorem for finite-dimensional quantum systems, based on our  discrete convolution, Theorem \ref{thm:CLT_gap}.
We also find a ``second law of thermodynamics for quantum convolution,'' Proposition \ref{prop;secondlaw}. This means that  quantum R\'enyi entropy $H_{\alpha}(\boxtimes^N\rho)$is non-decreasing with respect to the number $N$ of convolutions. Moreover,
 the repeated convolution of any zero-mean quantum state converges to the MS, with an exponential rate of convergence that is  bounded by the magic gap of the state, all stated precisely in Theorem \ref{thm:CLT_gap}. 
\end{enumerate}

In the case of CV quantum systems,  central limit theorems have  an interesting history that goes back to  Cushen and Hudson \cite{Cushen71}, and related work of Hepp and Lieb \cite{Lieb73,Lieb1973}. 
Many other quantum or noncommutative versions of the central limit theorem appeared later, see~\cite{Giri78,Goderis89,Matsui02,Cramer10,Jaksic09,Arous13,Michoel04,GoderisPTRT89,JaksicJMP10,Accardi94,Liu16,JiangLiuWu19,Hayashi09,CampbellPRA13,BekerCMP21,Carbone22}. 

For example,
in free probability theory
Voiculescu introduced and studied free convolution and proved a free central limit theorem:  
the repeated, normalized (additive) free convolution of a probability measure (with some assumptions) converges to 
a semicircle distribution~\cite{voiculescu1986addition,voiculescu1987multiplication,VDN92,voiculescu2016free}. 
The semicircle distribution in free probability plays a role similar to the Gaussian 
distribution in classical probability theory.

Several additional central limit theorems have been established in other frameworks. These include results for subfactor theory~\cite{Liu16,JiangLiuWu19}, for quantum walks on a lattice~\cite{Carbone22}, and for CV quantum information theory~\cite{CampbellPRA13,BekerCMP21}.

\section{Preliminaries}
We focus on the $n$-qudit system $\CHH^{\ot n}$, where $\CHH \simeq \complex^d$ is a $d$-dimensional Hilbert space and $d$ is any natural number. 
Let $D(\mathcal{H}^{\ot n})$ denote the set of  all quantum states on $\CHH^{\ot n}$.
In the Hilbert space $\mathcal{H}$, we consider the orthonormal, computational basis $\set{\ket{k}}_{k\in \mathbb{Z}_d}$. The Pauli $X$ and $Z$ operators
are 
\[ X: |k\rangle\mapsto |k+1\rangle,\;\;\; Z: |k\rangle \mapsto\xi^k_d|k\rangle,\;\;\;\forall k\in \mathbb{Z}_d\;, 
\]
where  $\mathbb{Z}_{d}$ is the cyclic group over $d$,  and $\xi_d=\exp(2\pi i /d)$ is a $d$-th root of unity. 
In order to define our quantum convolution, one needs to restrict $d$ to be prime.
If  $d$ is an odd prime number,  the local Weyl operators (or generalized Pauli operators)
are defined as 
$
w(p,q)=\xi^{-2^{-1}pq}_d\, Z^pX^q
$. Here $2^{-1}$ denotes the inverse $\frac{d+1}{2}$ of 2 in $\mathbb{Z}_d$.
If $d=2$, the Weyl operators are defined as 
$
w(p,q)=i^{-pq}Z^pX^q
$. Weyl operators for general local dimension $d$ are given in~\cite{Jaffe17}.  In the $n$-qudit system, the Weyl operators are defined as
 \begin{eqnarray*}
w(\vec p, \vec q)
=w(p_1, q_1)\ot...\ot w(p_n, q_n),
\end{eqnarray*}
with $\vec p=(p_1, p_2,..., p_n)\in \mathbb{Z}^n_d$, $\vec q=(q_1,..., q_n)\in \mathbb{Z}^n_d $,  which forms an orthonormal basis 
 with respect to the inner product 
$\inner{A}{B}=\frac{1}{d^n}\Tr{A^\dag B}$.  Denote $V^n:=\mathbb{Z}^n_d\times \mathbb{Z}^n_d$; this represents the phase space for $n$-qudit systems~\cite{Gross06}.
\begin{Def}
For any $n$-qudit state $\rho$, its characteristic function $\Xi_{\rho}:V^{n}\to\complex$ is
\begin{eqnarray*}
\Xi_{\rho}(\vec{p},\vec q):=\Tr{\rho w(-\vec{p},-\vec q)}.
\end{eqnarray*}
\end{Def}
Hence, 
any quantum state $\rho$ can be written as a linear combination of the Weyl operators 
\begin{eqnarray}\label{0109shi5}
\rho=\frac{1}{d^n}
\sum_{(\vec{p},\vec q)\in V^n}
\Xi_{\rho}(\vec{p},\vec q)w(\vec{p},\vec q).
\end{eqnarray}
The process of taking characteristic functions is the quantum Fourier transform that we consider. 
The characteristic function has been used to study quantum Boolean functions~\cite{montanaro2010quantum}. 
(See also a more general framework of quantum Fourier analysis \cite{JaffePNAS20}.)
The Clifford unitaries on $n$ qudits are the unitaries that map Weyl operators to  Weyl operators. 
Pure stabilizer states are pure states of the form $U\ket{0}^{\ot n}$, where $U$ is some Clifford unitary.
Equivalently, pure stabilizer states are the common eigenstates of  an abelian subgroup of the 
Weyl operators with size $d^n$. 
In general, let us consider any abelian subgroup of the Weyl operators 
with $r(\le n)$ generators $\set{w(\vec{p}_i, \vec q_i)}_{i\in[r]}$, 
and $[r]$ denotes the set $\set{0,1,2,...,r}$. 

\begin{Def}
A quantum state $\rho$ is a  {minimal stabilizer-projection state} (MSPS) associated with an abelian subgroup generated by $\set{w(\vec{p}_i, \vec q_i)}_{i\in[r]}$, if it has the following form 
\begin{eqnarray*}
\rho=
\frac{1}{d^{n-r}}\Pi^r_{i=1}\mathbb{E}_{k_i\in \mathbb{Z}_d}[\xi^{x_i}_dw(\vec{p}_i, \vec q_i)]^{k_i}\;,
\end{eqnarray*}
for some $(x_1,...,x_r)\in \mathbb{Z}^r_d$, with $\mathbb{E}_{k_i\in \mathbb{Z}_d}(\ \cdot \ ):=\frac{1}{d}\sum_{k_i\in\mathbb{Z}_d}(\ \cdot \ )$.
\end{Def}

An equivalent, alternative  definition is provided in the companion paper~\cite{BGJ23b}.
Let us consider an example with the abelian group $S=\set{Z_1,...,Z_{n-1}}$ for an $n$-qudit system.
The states $\set{\frac{1}{d}\proj{\vec j}\ot I}_{\vec j\in\mathbb{Z}^{n-1}_d}$ are MSPS. 
Moreover, a quantum state $\rho$ is called a stabilizer state if it can be written as a  convex combination of pure  
stabilizer states.

\section{Mean state}

In this section, we  introduce the notion of \textit{mean state} for a  given quantum state.
\begin{Def}[\bf Mean state (MS)]\label{def:mean_state}
Given an $n$-qudit state $\rho$,  the mean state  $\CMM(\rho)$ is the 
operator with the characteristic function: 
\begin{align}\label{0109shi6}
\Xi_{\CMM(\rho)}(\vec p, \vec q) :=
\left\{
\begin{aligned}
&\Xi_\rho ( \vec p, \vec q) , && |\Xi_\rho ( \vec p, \vec q)|=1,\\
& 0 , && |\Xi_\rho (  \vec p, \vec q)|<1.
\end{aligned}
\right.
\end{align}
 The mean state $\CMM(\rho)$ is an MSPS.
\end{Def}
We call $\CMM(\rho)$ the mean state because we use it to define 
the mean-value vector of the state $\rho$ in  \eqref{eq:mean_value} and the zero-mean state in Definition \ref{Def:Zero_mean}.
Moreover, we find that the MS is the closest MSPS in quantum R\'enyi relative entropy $D_{\alpha}$,
where 
\begin{eqnarray*}
    D_{\alpha}(\rho||\sigma):=\frac{1}{\alpha-1}\log\Tr{\left(\sigma^{\frac{1-\alpha}{2\alpha}}\rho\sigma^{\frac{1-\alpha}{2\alpha}}\right)^{\alpha}},
\end{eqnarray*}
and the quantum R\'enyi entropy is
\begin{eqnarray*}
H_{\alpha}(\rho)
:=\frac{1}{1-\alpha}\log \Tr{\rho^\alpha}, 
\end{eqnarray*}
for any $\alpha\in[0,+\infty]$. For example, the relative entropy $D(\rho||\sigma)=\lim_{\alpha\to 1}D_{\alpha}(\rho||\sigma)$, and the von Neuman
entropy $H(\rho)=\lim_{\alpha\to1}H_{\alpha}(\rho)$.

\begin{thm}[{\bf Extremality of MSPS}]\label{thm:main1}
Given an $n$-qudit state $\rho$ and $\alpha\in [1,+\infty]$, one has  
\begin{align*}
\min_{\sigma\in MSPS}D_{\alpha}(\rho||\sigma)
=D_{\alpha}(\rho||\CMM(\rho))=H_{\alpha}(\CMM(\rho))-H_{\alpha}(\rho).
\end{align*}
Moreover, $\CMM(\rho)$ is the unqiue minimizer, i.e., for any $\sigma\in MSPS$ with $\sigma\neq \CMM(\rho)$,  we have 
\begin{align*}
D_{\alpha}(\rho||\sigma)>D_{\alpha}(\rho||\CMM(\rho)).
\end{align*}
\end{thm}
Based on the above result, we can rewrite the quantum R\'enyi   entropy as follows
\begin{eqnarray}
H_{\alpha}(\CMM(\rho))=H_{\alpha}(\rho)+D_{\alpha}(\rho||\CMM(\rho)).
\end{eqnarray}
This equation shows the extremality of MSPS with respect to quantum R\'enyi  entropy:
within all quantum states having the same  MS up to Clifford conjugation, the MSPS $\CMM(\rho)$ attains the maximal value for  quantum R\'enyi  entropy, 
which we call "maximal entropy principle in DV systems.'' 
Recall the extremality of Gaussian states in CV systems, i.e.,  within all states
having a given covariance matrix,  Gaussian states attain the maximum
von Neumann entropy  \cite{Holevo99,WolfPRL06}. 
Hence the above theorem is the discrete version of the extremality of Gaussian states with the same covariance matrix in CV systems. 

In this work, 
we consider extremality properties of stabilizer states for quantum entropy. One can also consider the classical representation of quantum
states, for example by studying the characteristic functions.  One entropic measure, such as 
the $0$-R\'enyi-quantum-Fourier entropy of a pure state $\rho$ (defined as  the logarithm of the Pauli rank  $R_P(\rho)=|\text{Supp}(\Xi_{\rho})|$ ) also 
achieves its minimal value, iff $\rho$ is a stabilizer state~\cite{Bu19}. 
Other literature also touches on  classical descriptions of quantum states; for example extremality of pure coherent states in the Wehrl entropy is known, as are some variants~\cite{Weyl79,Lieb78,Carlen91,Lieb14}.

\begin{cor}\label{rem:RD}
In the resource theory of 
magic with $MSPS$  being the set of free states,
 the map from quantum states to $MSPS$, namely  $\rho\to\CMM(\rho)$, provides a nontrivial,  resource-destroying map. 
\end{cor}
Note that a map $\lambda$ from states to states is called a resource-destroying map \cite{LiuZiwenPRL17} if it satisfies two conditions: i) it maps all
quantum states to free states, i.e.~$\lambda(\rho)\in \mathcal{F}$ for any quantum state $\rho$, where $\mathcal{F}$ is the set of free states; 
ii) it preserves free states,
i.e., $\lambda(\sigma)=\sigma$ for any state $\sigma\in\mathcal{F}$.  The natural resource-destroying maps  are known in resource theories such as coherence, asymmetry, and non-Gaussianity (see Table \ref{tab:example}). 
However, it was unknown what a nontrivial, resource-destroying map is in the resource theory of magic. 
Here, our work shows that, the map $\CMM: \mathcal{D}(\mathcal{H}^{\ot n})\to MSPS$  is a resource-destroying map, which satisfies 
$\min_{\sigma\in MSPS}D_{\alpha}(\rho||\sigma)=D_{\alpha}(\rho||\mathcal{M}(\rho))$. 

\begin{table*}[htbp]
\begin{tabularx}{\textwidth}{lp{0.8\textwidth}}
Theory & Resource-destroying map\\  \hline\hline
Coherence &  $\Delta(\rho) = \sum_i\bra{i}\rho\ket{i}\proj{i}$, where $\Delta$  is the complete dephasing channel:$\{\ket{i}\}$ w.r.t. the  reference basis \cite{BaumgratzPRL14,Plenio17}.\\
Asymmetry & $\mathcal{G}(\rho) = \int_{G} d\mu(U) U\rho U^\dag$, where the integral is taken over the Haar measure on $G$  \cite{Gour09}. \\ 
Non-Gaussianity& 
$\lambda(\rho) = \rho_G$, where $\rho_G$ is the Gaussian state with the same mean displacement and covariance matrix as $\rho$ \cite{Marian13}.\\
Magic &  $\mathcal{M}(\rho)$, the closest MSPS  (Theorem \ref{thm:main1} in this work).\\\hline \hline 
\end{tabularx}
\caption{\label{tab:example}Resource theories with a nontrivial, resource-destroying map.}

\end{table*}

Since every quantum state $\rho$ can be written as a linear combination of the Weyl operators together with 
the characteristic function $\Xi_{\rho}$, the information of the state is encoded in the characteristic function.
We consider the gap between the largest absolute value, namely  1,  and the second-largest absolute value in the 
support of the characteristic function.
We call  
this the magic gap (or non-stabilizer gap).

\begin{Def}[\bf Magic gap]\label{def:ma_gap}
Given an $n$-qudit state $\rho\in\mathcal{D}(\mathcal{H}^{\ot n})$ for any integer $d\geq 2$, the  magic gap  of $\rho$ is 
\begin{eqnarray*}
MG(\rho)=1-\max_{(\vec{p}, \vec q)\in  \text{Supp}(\Xi_{\rho}): |\Xi_{\rho}(\vec{p},\vec q)|\neq 1}|\Xi_{\rho}(\vec{p},\vec q)|\;.
\end{eqnarray*}
If $\set{(\vec{p},\vec q)\in  \text{Supp}(\Xi_{\rho}): |\Xi_{\rho}(\vec{p}, \vec q)|\neq 1}=\emptyset$, define  $MG(\rho)=0$, i.e., there is no gap on the support of the characteristic function.
\end{Def}

\begin{prop}\label{prop:magap}
The magic gap (MG) of a state $\rho$ satisfies the following properties: 
\begin{enumerate}[]
\item{}
 The $MG(\rho)=0$, iff $\rho$ is an MSPS.
 Also
$0\leq MG(\rho)\leq 1-\sqrt{ \frac{d^n\Tr{\rho^2}-d^k}{R_P(\rho)-d^k}}$, where $R_P(\rho)=|\text{Supp}(\Xi_{\rho})|$ 
is the Pauli rank \cite{Bu19}.

\item{}
The $MG$ is invariant under Clifford unitaries. 

\item{} 
$MG(\rho_1\ot\rho_2)=\min\set{MG(\rho_1),MG(\rho_2)}$.
\end{enumerate}
\end{prop}

Since $-\log(1-x)= x+O(x^2)$, 
we can also consider the logarithmic magic gap (LMG), that is, 
\begin{eqnarray*}
LMG(\rho)=-\log\max_{(\vec{p},\vec q)\in  \text{Supp}(\Xi_{\rho}): |\Xi_{\rho}(\vec{p}, \vec q)|\neq 1}|\Xi_{\rho}(\vec{p}, \vec q)|.
\end{eqnarray*}
This $LMG(\rho)$ also satisfies conditions (1-3) in Proposition \ref{prop:magap} by changing the upper bound in (1) to 
$\frac{1}{2}\log\left[\frac{R_P(\rho)-d^k}{d^n\Tr{\rho^2}-d^k}\right]$.

Now, let us consider the application of the magic gap in the unitary synthesis. 
In an $n$-qubit system, 
the universal quantum 
circuits consist of  Clifford gates and $T$ gates.  From the Gottesman-Knill theorem~\cite{gottesman1998heisenberg}, we infer that Clifford unitaries  can be simulated efficiently on 
a classical computer. So  the $T$ gates (or other non-Clifford gates) are the source of any 
quantum computational advantage. Hence, it is important to determine how many $T$ gates are necessary to generate the target unitary. 
We find that the logarithm of the magic gap can  provide a lower bound on the number of $T$ gates.

\begin{prop}\label{prop:gap_syn}
Given an  input state $\rho$ and a quantum circuit $V_N$,  consisting of Clifford unitaries and $N$ magic T gates, 
the log magic gap of the output state $V_N\rho V^\dag_N$ satisfies, 
\begin{eqnarray*}
LMG(V_N\rho V^\dag_N)\leq LMG(\rho)+\frac{N}{2}.
\end{eqnarray*}
\end{prop}

\section{Convolution in DV quantum system}

We  introduce the convolution between 2 different $n$-qudit systems,  denoted by  
$\mathcal{H}_A$ and $\mathcal{H}_B$, respectively. In other words, the Hilbert spaces are $\mathcal{H}_A=\mathcal{H}^{\ot n}$, and 
$\mathcal{H}_B=\mathcal{H}^{\ot n}$, where dim $\mathcal{H}=d$.

\subsection{Discrete convolution}
 Given a prime number $d$, consider the $2\times 2 $ invertible  matrix of parameters,
\begin{equation}\label{0125shi1}
G=\left[
\begin{array}{cc}
g_{00}&g_{01}\\
g_{10}&g_{11}
\end{array}
\right]:=[g_{00},g_{01};g_{10},g_{11}]\;,
\end{equation}
with entries in $\mathbb{Z}_d$. 
We assume that $G$ is invertible in $\mathbb{Z}_{d}$, so  $\det \,G=g_{00}g_{11}-g_{01}g_{10} \not\equiv 0\mod d$. The inverse  in $\mathbb{Z}_d$ is  
\begin{equation}
G^{-1}=N\left[
\begin{array}{cc}
\phantom{-}g_{11}&-g_{01}\\
-g_{10}&\phantom{-}g_{00}
\end{array}
\right], \quad
\text{where } N=(\det \,G)^{-1}\;.
\end{equation} 
The  matrix $G$ is called \textit{positive} if none of 
$g_{ij}\equiv 0 \mod d$. In this work, we focus on the case 
where $G$ is positive and invertible. 
If $d$ is an odd prime number, 
there always exists a positive and invertible matrix $G$ in $\mathbb{Z}_d$, e.g., 
$G=[1,1;1,d-1]$. If $d=2$,  there is no positive and  invertible matrix $G$, as the only positive
matrix $[1,1;1,1]$ is not invertible in $\mathbb{Z}_2$.

\begin{Def}[\bf Key unitary]\label{def:Cli_unitary}
Given a positive and invertible matrix $G$,  a $2n$-qudit unitary $U$ is
\begin{eqnarray}\label{eq:Cli_unitary}
U=\sum_{\vec{i},\vec j}\ket{\vec{i}'}\bra{\vec i}\ot \ket{\vec{j}'}\bra{\vec j} \;,
\end{eqnarray}
 where the state $| \vec i \rangle = |  i_1 \rangle \otimes \cdots \otimes |  i_n \rangle \in \CHH^{\otimes n} $,
 and $\left[\begin{array}{c}
 i'_k\\
 j'_k
 \end{array}\right]=(G^{-1})^T
 \left[\begin{array}{c}
 i_k\\
 j_k
 \end{array}
 \right]
 $, for  $k\in [n]$.
\end{Def}
That is, $U$ maps the state $| \vec i , \vec j \rangle $ to the state $ \ket{Ng_{11}\vec{i}-Ng_{10}\vec{j}, -Ng_{01}\vec{i}+Ng_{00}\vec{j}}$,  where $N=(det\,G)^{-1}=(g_{00}g_{11}-g_{01}g_{10})^{-1}$.

\begin{Def}[\bf Convolution of states]\label{def:convo}
Given the Clifford unitary $U$ in \eqref{eq:Cli_unitary}, and two quantum states $\rho\in \mathcal{D}(\mathcal{H}_A),\sigma\in \mathcal{D}(\mathcal{H}_B)$,
the convolution  of $\rho$ and $\sigma$ is  
\begin{align}\label{eq:convo}
\rho \boxtimes \sigma = \ptr{B}{ U (\rho \otimes \sigma) U^\dag}.
\end{align}
The partial trace is taken on the second $n$-qudit system $\mathcal{H}_B$.
The corresponding quantum convolutional channel $\mathcal{E}$ is
\begin{eqnarray}\label{eq:convo_chn}
\mathcal{E}(\rho_{AB} )=\ptr{B}{ U( \rho_{AB}) U^\dag},
\end{eqnarray}
for any quantum state $\rho_{AB}$ on $\mathcal{H}_{A}\ot\mathcal{H}_{B}$.
\end{Def}

\begin{prop}[{\bf Convolution-multiplication duality}]
Given the convolution with the parameter matrix $G$, then the characteristic function  satisfies 
\begin{align*}
\Xi_{ \rho \boxtimes \sigma} (\vec p, \vec q) = \Xi_\rho (Ng_{11}\vec p, g_{00}\vec q)\; \Xi_\sigma (-Ng_{10}\vec p, g_{01}\vec q)\;,
\end{align*}
for any $\vec p, \vec q \in \mathbb{Z}^n_d$.
\end{prop}

In classical probability theory, the convolution of 
two Gaussian distributions is still a Gaussian. Here, we find the analogous 
property for stabilizer states.
\begin{prop}[{\bf Convolutional stability}] \label{prop:convolutionalstability}
Given two $n$-qudit stabilizer states $\rho$ and $\sigma$, $\rho\boxtimes\sigma$ is a stabilizer state.
\end{prop}
It is well-known that the distance measure is monotone under the convolution $*$ in 
classical probability theory,
$$
D(\mu_1*\nu,\mu_2*\nu)\leq D(\mu_1,\mu_2)\;,
$$
for measures $\mu_1,\mu_2,\nu$  on $\real^d$,
where $D$ is either the classical total variation distance, relative divergence, or Wasserstein distance. 
Here we establish a quantum version of the monotonicity of distance measures under quantum convolution, for the distance measures including the
$L_1$ norm, relative entropy, and quantum Wasserstein distance (defined in \cite{Palma21}). 

\begin{prop}[{\bf Monotonicity under convolution}]
Let the distance measure $D:\mathcal{D}(\mathcal{H}^{\ot n})\times \mathcal{D}(\mathcal{H}^{\ot n})\to \real$  be the $L_1$ norm, relative entropy, or  quantum Wasserstein distance.  Then for any convolution $\boxtimes$ with respect to the positive and invertible matrix $G$, we have 
\begin{eqnarray}
D(\rho\boxtimes\tau, \sigma\boxtimes\tau)
\leq D(\rho, \sigma).
\end{eqnarray}
\end{prop}

\subsection{Quantum entropy and Fisher-information inequalities }
Consider the behavior of the generalized quantum R\'enyi entropy \cite{BHOW15}
under convolution. Here 
\begin{eqnarray}
H_{\alpha}(\rho)
:=\frac{\sgn(\alpha)}{1-\alpha}\log {\sum_i\lambda^{\alpha}_i}\;,
\quad \forall \alpha\in [-\infty, +\infty],
\end{eqnarray}
where $\lambda_i$ are the eigenvalues of $\rho$, and $\sgn(\alpha)=\pm 1$.

\begin{thm}[\bf Convolution increases R\'enyi entropy]\label{thm:entropy}
Let the parameter matrix $G$ be positive and invertible, and $\rho, \sigma$ be two $n$-qudit states.
The generalized R\'enyi entropy satisfies
\begin{eqnarray}
H_{\alpha}(\rho\boxtimes\sigma)
\geq \max\set{H_{\alpha}(\rho), H_{\alpha}(\sigma)}\;,
\end{eqnarray}
for any $\alpha\in[-\infty,+\infty]$.
\end{thm}

Besides quantum R\'enyi entropies, we also consider the divergence-based quantum Fisher information \cite{Konig14}:
given a smooth one-parameter family of states $\set{\rho_{\theta}}_{\theta}$, the divergence-based quantum Fisher information at $0$
is defined as 
\begin{eqnarray*}
J(\rho^{\theta};\theta)\big|_{\theta=0}
:=\frac{d^2}{d\theta^2}\bigg|_{\theta=0}D(\rho||\rho_{\theta}).
\end{eqnarray*}
Since the first derivative $\frac{d}{d\theta} \big|_{\theta=0}D(\rho||\rho_{\theta})=0$, 
the second derivative $J(\rho^{\theta};\theta)|_{\theta=0}$
quantifies the sensitivity of the divergence with respect to the change of parameter $\theta$.
Since we only consider the divergence-based quantum Fisher information $J(\rho^{\theta};\theta)|_{\theta=0}$ in this work, 
we call it the quantum Fisher information for simplicity. 
If $\set{\rho_{\theta}}_{\theta}$ is a family of parameterized states defined by 
$\rho_{\theta}=\exp(i\theta H)\rho\exp(-i\theta H)$ with respect to a Hermitian operator $H$ for all $\theta\in \real$ , then 
the quantum Fisher information can be written as
\begin{align*}
J( \rho; H )=\frac{d^2}{d\theta^2} \bigg|_{\theta=0} D(\rho||\rho_{\theta}) = \Tr{ \rho [H, [H, \log \rho]]}.
\end{align*}

In $n$-qudit systems,
we denote $X_k$ (resp., $Z_k$) to be the Pauli $X$ (resp., $Z$) operator on $k$-th qudit.
For $R= X_k$ or $Z_k$ ($1\le k\le n$),
denote $|j\rangle_R $ to be an eigenvector of $R$  corresponding to the eigenvalue $\xi^j_d$ with $j\in \mathbb{Z}_d$.
Let us define the Hermitian operator  $H^R_j$ for $j\in [d]$ as 
$
H_j^R = \proj{j}_R
$,
and the corresponding parameterized unitary $U_j^R(\theta)$ as
$
U_j^R(\theta) = \exp( i \theta H_j^R )
$. 
Then, for any quantum state $\rho$, 
let us consider the family of parameterized states
$ \rho_{R,\theta}=U_j^R(\theta)\rho U_j^R(\theta)^\dag, \theta\in\real$, 
and the corresponding quantum Fisher information 
$J(\rho; H^R_j)$. 
Let us denote 
\begin{align}\label{0103shi6}
J(\rho) = \sum_{k=1}^n \sum_{j=1}^d J(\rho; H_j^{X_k}) + J(\rho; H_j^{Z_k}).
\end{align}

\begin{thm}[\bf Convolution decreases Fisher information]\label{thm:fisher}
Let the parameter matrix $G$ be positive and invertible, and $\rho, \sigma$ be two $n$-qudit states.
The quantum Fisher information satisfies
\begin{align}
J( \rho\boxtimes \sigma) \le \min\set{J(\rho), J(\sigma)}.
\end{align}

\end{thm}

\subsection{Stabilizer states in the convolutional channel}

What kind of input states $\rho,\sigma$ will make the output state have the minimal output entropy?

\begin{thm}\label{thm:min_outen}
Let the parameter matrix $G$ be positive and invertible, and $\rho, \sigma$ be two $n$-qudit states.
The output state $\mathcal{E}(\rho\ot\sigma)$ has the minimal 
output entropy iff both $\rho$ and $\sigma$ are pure stabilizer states,
and the stabilizer groups $S_1$ and $S_2$ of $\rho$ and $\sigma$ satisfy 
\begin{align}\label{0125shi2}
S_1 = \{ w(- g_{10}^{-1} g_{11}\vec x, g_{01}^{-1} g_{00}\vec y):   w(\vec x, \vec y)\in S_2\}.
\end{align}
\end{thm}

Besides, we consider the 
Holevo capacity of the quantum channel, which can be used to quantify  
the classical capacity of a memoryless quantum channel \cite{Schumacher97,Holevo98}. 
\begin{Def}[\bf Holevo capacity]
Given a quantum channel $\mathcal{E}$, 
the Holevo capacity $\chi(\mathcal{E})$ is 
\begin{eqnarray*}
\chi(\mathcal{E})
=\max_{\set{p_i,\rho_i}}
H\left(\sum_ip_i\mathcal{E}(\rho_i)\right)
-\sum_ip_iH(\mathcal{E}(\rho_i)),
\end{eqnarray*}
where the maximum is taken over all ensembles $\set{p_i,\rho_i}$ of possible input states $\rho_i$ occurring with
probabilities $p_i$.
\end{Def}

Given a quantum state $\sigma$, the quantum chanenl $\mathcal{E}_{\sigma}(\cdot)=\mathcal{E}(\cdot\ot \sigma)$. That is, 
 for any input state $\rho$,
the output state of the channel $\mathcal{E}_{\sigma}$ is $\rho\boxtimes \sigma$. We find that 
the Holevo capacity of  $\mathcal{E}_{\sigma}$ can be bounded by the 
 entropies of both $\sigma$ and $\CMM(\sigma)$.

\begin{thm}[\bf Holevo capacity bound: general case]\label{0118thm1}
Let the parameter matrix $G$ be positive and invertible, and $\sigma$ be an $n$-qudit state.
The Holevo capacity of the quantum channel $\mathcal{E}_{\sigma}$ is 
\begin{eqnarray}\label{eq:lo_up_hol}
n\log d-H(\CMM(\sigma))\leq \chi(\mathcal{E}_{\sigma})\leq  n\log d-H(\sigma).
\end{eqnarray}
If $\sigma\in MSPS$,
then 
\begin{eqnarray}
\chi(\mathcal{E}_{\sigma})
=n\log d-H(\sigma).
\end{eqnarray}
\end{thm}

Besides, we find that the pure stabilizer states are the only states making the 
convolutional channel $\mathcal{E}_{\sigma}$ achieve the maximal Holevo capacity.
\begin{thm}[\bf Maximizer for Holevo capacity: pure stabilizer states]\label{thm:holv_stab}
Let the parameter matrix $G$ be positive and invertible.
The quantum channel $\mathcal{E}_{\sigma}$  has the maximal Holevo capacity
$n\log d$ iff 
$\sigma$ is a pure stabilizer state.
\end{thm}

\subsection{Examples: discrete beam splitter and amplifier}\label{subsec:examp}

Now, let us consider two examples of  convolutions. The first one is 
the discrete beam splitter with  $G=[s,t;t,-s]$
and $s^2+t^2\equiv 1\mod d$.
This  is a discrete version of the condition 
$(\sqrt{\lambda})^2+(\sqrt{1-\lambda})^2=1$ that occurs in CV  beam splitter.  In fact, 
the condition $s^2+t^2\equiv 1 \mod d$ can be satisfied for any prime number $d\ge 7$ by some number theory guarantee. 
Formally, we have the following definition of discrete splitter beam.
\begin{Def}[\bf Discrete beam splitter]\label{Def:disc_BS}
Given $s^2+t^2\equiv 1 \mod d$,  the unitary operator $U_{s,t}$ is
\begin{align}\label{1231shi1}
 U_{s,t} = \sum_{\vec i,\vec j\in \mathbb{Z}^n_d} |s\vec i+t\vec j \rangle \langle \vec i| \otimes | t\vec i-s\vec j\rangle \langle \vec j|,
 \end{align}
 where the state$| \vec i \rangle = |  i_1 \rangle \otimes \cdots \otimes |  i_n \rangle \in \CHH^{\otimes n} $.
The convolution  of two $n$-qudit states $\rho$ and $\sigma$ is 
\begin{align}\label{eq:conv_B}
\rho \boxtimes_{s,t} \sigma = \ptr{B}{ U_{s,t} (\rho \otimes \sigma) U^\dag_{s,t}}.
\end{align}
\end{Def}
 We summarize  and  compare our results on 
 discrete beam splitter  with the known results for CV quantum systems in Table~\ref{tab:sum_B}.

\begin{table*}[!htbp]
\centering
\begin{tabular}{ |c|c|c|c| } 
\hline
Beam splitter& CV quantum systems & DV quantum systems\\
\hline
Parameter& $(\sqrt{\lambda},\sqrt{1-\lambda}),\lambda\in[0,1]$ & $(s,t)$, $s^2+t^2\equiv1 \mod d$ \\
\hline
\multirow{2}{*}{Convolution}& 
$\rho\boxtimes_{\lambda}\sigma=\ptr{B}{U_{\lambda}\rho\ot\sigma U^\dag_{\lambda}}$, & $\rho\boxtimes_{s,t}\sigma=\ptr{B}{U_{s,t}\rho\ot\sigma U^\dag_{s,t}}$,\\
& $U_{\lambda}$: beam splitter  & $ U_{s,t}$ : discrete beam splitter  \\
\hline
Characteristic function &  $\Xi_{\rho\boxtimes_{\lambda}\sigma}(\vec{x})=\Xi_{\rho}(\sqrt{\lambda}\vec{x})\Xi_{\sigma}(\sqrt{1-\lambda}\vec{x})$&$\Xi_{\rho\boxtimes_{s,t}\sigma}(\vec{x})=\Xi_{\rho}(s\vec{x})\Xi_{\sigma}(t\vec{x})$\\
\hline
\multirow{2}{*}{Quantum entropy inequality}& $H(\rho\boxtimes_{\lambda}\sigma)\geq \lambda H(\rho)+(1-\lambda)H(\sigma)$ \cite{Konig14} &
$H_{\alpha}(\rho\boxtimes_{s,t}\sigma)\geq \max\set{H_{\alpha}(\rho),H_{\alpha}(\sigma)}$,\\
&$e^{H(\rho\boxtimes_{\lambda}\sigma)/n}\geq \lambda e^{H(\rho)/n}+(1-\lambda)e^{H(\sigma)/n}$ \cite{Konig14,Palma14}
& $\alpha\in[-\infty,\infty]$(Theorem \ref{thm:entropy})\\
\hline 
\multirow{2}{*}{Quantum Fisher information inequality}&
$w^2J(\rho\boxtimes_{\lambda}\sigma)\leq w^2_1J(\rho)+w^2_2J(\sigma)$,&
$J(\rho\boxtimes_{s,t}\sigma)\leq \min\set{J(\rho),J(\sigma)}$\\
& $w=\sqrt{\lambda}w_1+\sqrt{1-\lambda}w_2$ \cite{Konig14}&(Theorem \ref{thm:fisher})\\ 
\hline
\end{tabular}
\caption{\label{tab:sum_B}Comparison of results for the CV and DV beam splitters.}

\end{table*}

Besides the discrete beam splitter, we  define the discrete amplifier
with $G=[l,-m;-m,l]$
and  $l^2-m^2\equiv 1\mod d$.
This  is a discrete version of the condition 
$(\sqrt{\kappa})^2+(\sqrt{\kappa-1})^2=1$ with $\kappa\in[1,\infty]$ that occurs in CV squeezing unitary.  In fact, 
the condition $l^2-m^2\equiv 1\mod d$ can also be satisfied for any prime number $d\ge 7$ by some number theory guarantee. 
Formally, we have the following definition of discrete  amplifier. 

\begin{Def}[\bf Discrete  amplifier]\label{Def:dis_sq}
Given $l^2-m^2\equiv 1 \mod d$, 
the unitary operator $V_{l,m}$ is
\begin{align}\label{eq:squeez}
 V_{l,m} = \sum_{\vec i,\vec j\in \mathbb{Z}^n_d} |l\vec i+m\vec j \rangle \langle \vec i| \otimes | m\vec i+l\vec j\rangle \langle \vec j|.
 \end{align}
 The convolution of two $n$-qudit states $\rho$ and $\sigma$ 
is 
\begin{align}\label{eq:conv_S}
\rho \boxtimes_{l,m} \sigma = \ptr{B}{ V_{l,m} (\rho \otimes \sigma) V^\dag_{l,m}}.
\end{align}
\end{Def}
We summarize  and  compare our results on 
 discrete  amplifier with the known results for CV quantum systems in Table~\ref{tab:sum_S}.

\begin{table*}[!htbp]
\centering
\begin{tabular}{ |c|c|c|c| } 
\hline
Amplifier & CV quantum systems & DV quantum systems\\
\hline
Parameter& $(\sqrt{\kappa},\sqrt{\kappa-1}),\kappa\in[1,\infty]$ & $(l,m)$, $l^2-m^2\equiv1 \mod d$ \\
\hline
\multirow{2}{*}{Convolution}& 
$\rho\boxtimes_{\kappa}\sigma=\ptr{B}{V_{\kappa}\rho\ot\sigma V^\dag_{\kappa}}$, & $\rho\boxtimes_{l,m}\sigma=\ptr{B}{V_{l,m}\rho\ot\sigma V^\dag_{l,m}}$,\\
& $V_{\kappa}$ : squeezing unitary & $V_{l,m}$ : discrete squeezing unitary \\
\hline
Characteristic function &  $\Xi_{\rho\boxtimes_{\kappa}\sigma}(\vec p, \vec q)=\Xi_{\rho}(\sqrt{\kappa}\vec p, \sqrt{\kappa} \vec q)\;\Xi_{\sigma}(\sqrt{\kappa-1}\vec p, -\sqrt{\kappa-1}\vec q)$&$\Xi_{\rho\boxtimes_{l,m}\sigma}(\vec{p},\vec q)=\Xi_{\rho}(l\vec{p}, l\vec q)\;\Xi_{\sigma}(m\vec p, -m\vec q)$\\
\hline
\multirow{2}{*}{Quantum entropy inequality}&$e^{H(\rho\boxtimes_{\kappa}\sigma)/n}\geq \kappa e^{H(\rho)/n}+(\kappa-1)e^{H(\sigma)/n}$  &
$H_{\alpha}(\rho\boxtimes_{l,m}\sigma)\geq \max\set{H_{\alpha}(\rho),H_{\alpha}(\sigma)}$,\\
& \cite{Palma14}
& $\alpha\in[-\infty,+\infty]$(Theorem \ref{thm:entropy})\\
\hline 
\multirow{2}{*}{Quantum Fisher information inequality}&
$w^2J(\rho\boxtimes_{\kappa}\sigma)\leq w^2_1J(\rho)+w^2_2J(\sigma)$,&
$J(\rho\boxtimes_{l,m}\sigma)\leq \min\set{J(\rho),J(\sigma)}$\\
& $w=\sqrt{\kappa}w_1+\sqrt{\kappa-1}w_2$ \cite{Palma14}&(Theorem \ref{thm:fisher})\\ 
\hline
\end{tabular}
\caption{\label{tab:sum_S}Comparison of results for the CV and DV amplifiers.}
\end{table*}

\section{The Central limit theorem}\label{sec:CLL}

Let us denote that $\boxtimes^{N+1}\rho=(\boxtimes^N\rho)\boxtimes\rho$, and $\boxtimes^0\rho=\rho$, where $\boxtimes$ is short for the beam splitter convolution $\boxtimes_{s,t}$ in \ref{1231shi1} (which does not require $s\equiv t\mod d$). 
By applying Theorem \ref{thm:entropy},
we find that quantum R\'enyi entropy $H_{\alpha}(\boxtimes^N\rho)$ is increasing w.r.t. the number of convolutions $N$.

\begin{prop}[\bf Second law of thermodynamics for quantum convolution]\label{prop;secondlaw}
For any $n$-qudit state $\rho$,
the quantum R\'enyi entropy  satisfies the following property,
\begin{eqnarray}
H_{\alpha}(\boxtimes^{N+1}\rho)
\geq H_{\alpha}(\boxtimes^N\rho)\;,\quad \forall N\geq 0\;,
\end{eqnarray}
for any $\alpha\in [-\infty, +\infty]$.
\end{prop}

Note that 
in the classical case \cite{artstein2004JAMS,Lieb78,McKean1966}, it was proved that
$
H\left(\frac{X_1+X_2+...X_{N+1}}{\sqrt{N+1}}\right)
\geq H\left(\frac{X_1+X_2+...X_{N}}{\sqrt{N}}\right)
$,
where  $X_1, X_2,...$ are  i.i.d., square-integrable random variables; this is 
a classical analogue of the second law of thermodynamics.

Before considering the quantum central limit theorem, let us first look at the classical 
case. 
Given a random variable $X$ with probability density function $f$, 
if $X$ has zero mean,
then  $\frac{1}{\sqrt N}X_1+\cdots + \frac{1}{\sqrt N}X_N$ will converge  to some normal 
random variable, that is, the  probability density function $ f_{\boxtimes N} $ converges to a normal distribution  as $N\rightarrow \infty$,
where $f_{\boxtimes N}$ denotes the balanced  $N$-th convolution of $f$. 
The condition that $X$ has zero mean cannot be removed.
For example,
if  $X\sim \CNN(1,1)$,
$\frac{1}{\sqrt N}X_1+\cdots + \frac{1}{\sqrt N}X_N\sim \CNN(\sqrt N, 1)$ and doesn't have a limit distribution.
Hence, given a random variable $X$, we should consider the zero-mean variable $X-\mathbb{E}X$ instead of $X$, where $\mathbb{E}X$ denotes the mean value of $X$.

For classical multi-variable random variable $\vec{X}\in \real^r$, %with mean vector $\vec{\mu}=(\mu_1,..,\mu_r)\in \real^r$, 
its characteristic function 
is
\begin{eqnarray*}
\phi_{X}(\vec t)=\mathbb{E}_{\vec{X}}\exp(i\vec{t}\cdot \vec{X}),\quad \forall \vec{t}=(t_1,..,t_r)\in \real^r,
\end{eqnarray*}
and mean-value vector $\vec \mu$ equals to the  gradient of $\phi_{X}(\vec t)$ at $\vec t= \vec 0$, i.e., 
\begin{eqnarray*}
\vec{\mu}=\left(-i\left.\frac{d}{d t_j}\right|_{\vec t=\vec 0}\phi_{X}(\vec t)\right)^r_{j=1}.
\end{eqnarray*}
If $X$ is zero-mean, then $\vec{\mu}=(0,...,0)$.

For the quantum case, we also need to define the 
zero-mean state to consider the quantum central limit theorem. 
Given an $n$-qudit state $\rho$,  the MS $\mathcal{M}(\rho)$ has the
characteristic function 
\begin{eqnarray*}
\Xi_{\mathcal{M}(\rho)}\left(\sum^r_{i=1}t_i(\vec{p}_i, \vec{q}_i)\right)
=\Pi^r_{i=1}\xi^{t_ik_i}_d=\xi^{\sum_it_ik_i}_d,
\end{eqnarray*}
where we assume the abelian group of $\CMM(\rho)$ is generated by the Weyl operators $\set{w(\vec{p}_i, \vec{q}_i)}_{i\in[r]}$, and 
$\Xi_{\rho}(\vec p_i, \vec q_i)=\xi^{k_i}_d, \forall i\in[r]$.
Similar to the classical case, we 
define 
the mean-value vector of the state $\rho$ w.r.t. the generators $\set{w(\vec{p}_i, \vec{q}_i)}_{i\in[r]}$ as 
\begin{eqnarray}\label{eq:mean_value}
\vec{\mu}_{\CMM(\rho)}
=(k_1,...,k_r) \mod d .
\end{eqnarray}

\begin{Def}[\bf Zero-mean state]\label{Def:Zero_mean}
Given an $n$-qudit state $\rho$, $\rho$ is called a zero-mean state if  $\mathcal{M}(\rho)$ has mean-value 
vector $\vec{\mu}_{\CMM(\rho)}=(0,..., 0)\mod d$, or equivalently 
the characteristic function of $\mathcal{M}(\rho)$ takes values in $\set{0, 1}$. 
\end{Def}
In fact,  if $\rho$ is not a zero-mean state, there exists a Weyl operator $w(\vec p, \vec q)$ such that $w(\vec p, \vec q)\rho w(\vec p, \vec q)^\dag$
is a zero-mean state.
Now, we have the following result on quantum central limit theorem  for  the
$L_2$ norm, where 
the rate of convergence is controlled by the 
 magic gap.

\begin{thm}[\bf Central limit theorem via magic gap]\label{thm:CLT_gap} 
Given a zero-mean  $n$-qudit state $\rho$, we have 
\begin{eqnarray}
\norm{\boxtimes^N\rho-\CMM(\rho)}_2
\leq (1-MG(\rho))^{N}\norm{\rho-\CMM(\rho)}_2.
\end{eqnarray}
If $\rho\neq \CMM(\rho)$, then  $MG(\rho)>0$, and the rate of convergence is exponentially small with respect to the time of
convolution.
\end{thm}

\section{Some open problems}
There are many open questions, such as:
\begin{enumerate}[]
\item{}
Aside from Clifford unitaries, matchgate~\cite{valiant2002quantum}, or Gaussian ferimonic operations~\cite{terhal2002classical,divincenzo2004fermionic,jozsa2008matchgates}, is another tractable family of quantum circuits. Could our convolution be helpful to define matchgates for qudits?

\item{}

 In graph theory, the Cheeger constant measures the edge expansion of a graph. The Cheeger inequalities relate the spectral gap of the adjacency matrix of a graph to its Cheeger constant \cite{Cheeger71, dodziuk1984difference, alon2016probabilistic}. Is there a quantum Cheeger constant that corresponds to the magic gap?
 
 \item{}
 Can one generalize our convolution using picture languages, such as the Quon language \cite{Jaffe17}, the tensor network \cite{Orus14,Biamonte17}, the ZX calculus \cite{Coecke08,Coecke11}, or so on?

\item{}
 Following the convolution proposed and studied in~\cite{Audenaert16, Carlen16}, many generalizations have been studied  \cite{Jeong2018,HuangLiuWu22}, 
including a generalization to certain von Neumann algebras~\cite{HuangLiuWu22}. Will similar generalizations  be possible  for the convolution in this paper? 

\item{}
Similar to the classical case~\cite{Barron86, Kontoyiannis21},
we can explore the entropic limit theorem for quantum convolution. Due to the continuity of relative entropy, $D(\boxtimes^N\rho||\CMM(\rho))$ converges to $0$. Can one determine the rate of convergence?

\item{}
Clarify the relation between the 
convolution and central limit theorem in this work to their counterparts in free probability theory. 
The free convolution corresponds to the free independence of random variables. What independence relation corresponds to our convolution?

 \end{enumerate}

 \section{Acknowledgments}
We thank Eric Carlen, Roy Garcia,  Jiange Li, Seth Lloyd, Sijie Luo for  helpful discussions. This work was supported in part by ARO Grant W911NF-19-1-0302 and ARO MURI Grant W911NF-20-1-0082, and NSF Eager Grant 2037687.

\bibliography{reference}{}

\end{document}